\documentclass[nofootinbib, reprint, preprintnumbers, amsmath, amssymb, aps, pra, superscriptaddress, floatfix]{revtex4-1}
\usepackage[utf8]{inputenc}
\usepackage{mathtools}
\usepackage{graphicx}
\usepackage{dcolumn}
\usepackage{bm}
\usepackage{physics}
\usepackage{bm}
\usepackage{color}
\usepackage{braket}

\usepackage{mathrsfs}
\usepackage{bbm}

\usepackage{comment}
\usepackage{lipsum}

\begin{document}
\preprint{KEK-TH-2325}

\title{Weak value amplification and the lifetime of decaying particle}
\author{Yuichiro Mori}
\email{yuichiro@post.kek.jp}
\affiliation{Theory Center, Institute of Particle and Nuclear Studies, High Energy Accelerator Research Organization (KEK), Ibaraki 305-0801, Japan}%
\author{Izumi Tsutsui}
\email{izumi.tsutsui@kek.jp}
\affiliation{Theory Center, Institute of Particle and Nuclear Studies, High Energy Accelerator Research Organization (KEK), Ibaraki 305-0801, Japan}%
\affiliation{Department of Physics, The University of Tokyo, Tokyo 113-0033, Japan}
\date{\today}

\begin{abstract}

We study the possibility of varying the measured lifetime of a decaying particle based on the technique of weak value amplification in which an additional filtering process called postselection is performed.  Our analysis made in a direct measurement scheme presented here 
shows that, for simple two-level systems, the lifetime may be prolonged more than three times compared to the original one, while it can also be shortened arbitrarily by a proper choice of postselection.  This result is consistent with our previous analysis on the possible prolongation of the lifetime of $B$ mesons that may be observed in laboratories, and suggests room for novel applications of weak value amplification beyond precision measurement conventionally considered. 

\end{abstract}

\pacs{\textcolor{blue}{Valid PACS appear here}}

\maketitle

\section{Introduction}\label{sec:introduction}

The discovery of unstable nuclei at the turn of the 20 century
was generally regarded as a primary source of the development of quantum theory,  along with the invention of energy quanta compelled by  blackbody radiation.  The lifetime of an unstable state or a decaying particle is 
a fundamental physical property inherent to the particle, which is believed to be firmly determined by nature.  
Nearly 90 years later, a novel notion of physical quantity called the weak value was proposed by Aharonov \textit{et al}.~\cite{Aharonov_1988}, which has the peculiar property of admitting \lq amplification\rq, allowing its measured values to vary from the original ones considerably.  For instance, the electronic spin, $\hbar/2$, can in principle be enlarged 100 times or more with the weak value, although the implementation of the measurement may impose restrictions on the amplification for practical reasons.

The weak value amplification can be achieved by taking a step called postselection, in which one chooses the final state in the physical process properly.  This method of amplification has an obvious advantage for precision measurement, and indeed we have seen successful applications in various fields including optical systems \cite{Hosten_2008, Dixon_2009, Li_2018}, neutron systems \cite{Denkmayr:2014aa} and cold atomic systems \cite{CWWU_2019}.  
One may then ask whether lifetime can also be put into the form of the weak value and, if so, how much the decay of an unstable particle can be delayed by the amplification procedure.  
Unfortunately, these preceding experiments were not suitable for extending the measured lifetime, given that the procedure to obtain the weak value proposed in \cite{Aharonov_1988} was based on the indirect measurement scheme using probes, while the recently proposed methods without using probes \cite{Lan_2019, Ogawa_2020} were not optimized for the measurement of lifetime.

An adequate procedure of weak value amplification for lifetime has been discussed by Shomroni \textit{et al}.~in 2013 \cite{Shomroni_2013} for a system exhibiting spontaneous emission employing the direct measurement scheme.  It was reported that the lifetime can effectively be made 1.9 times longer than the original lifetime 26 $\mathrm{ns}$ through postselection.  Recently, we have used their scheme for studying the possible precision measurement of the $CP$ parameter in the $B$ meson decay \cite{Higashino_2020}, where we found that the lifetime of the meson can be made 2.64 times longer than the original lifetime 1.5 $\mathrm{ps}$.  Naturally, these numbers of prolongation found in specific systems prompt us to find the general range of weak value amplification available in this scheme.  This is precisely the aim of the study presented in this paper.

More explicitly, here we provide a workable scheme of weak value amplification designed for lifetime in a general form, and thereby examine the range of amplification for systems possessing two energy levels to gain a benchmark for the lifetime prolongation in more general cases.  Our result shows that the amplification scheme can enlarge the lifetime more than three (up to 3.414) times.  Although this amplification factor is not radical, it could nevertheless be significant when one wants to further improve precision for lifetime measurement or one wishes to gain time to perform some manipulation before decay.

This paper is organized as follows.  After we briefly recall in Sec.~\ref{sec:time_distribution} the basics of time evolution of unstable particles,  we provide in Sec.~\ref{sec:time_distribution_post} the decay time distribution when the postselection is made.  Then, in Sec.~\ref{sec:lifetime_and_its_range}, we discuss the possible range of lifetime that can be extended or shortened as a result of the postselection, specializing into the two level systems where the closed form of lifetime can be available.  
Our conclusions are summarized in Sec.~\ref{sec:conclusion_and_discussion}.  Two appendices are provided to support the argument given in the text.

\section{Time evolution of unstable particles}\label{sec:time_distribution}

Following the standard formulation, we shall describe the time evolution of unstable particles phenomenologically by means of a non-Hermitian Hamiltonian $\hat{H}$.  When it is allowed to restrict ourselves to a pair of energy eigenstates of the particle system, the Hamiltonian 
may in general be characterized by the eigenvalue equations, 
\begin{align}
\hat{H}|P_{\mathrm{L}}\rangle&=\left(M_{\mathrm{L}}-\frac{i}{2}\Gamma_{\mathrm{L}}\right)|P_{\mathrm{L}}\rangle,\label{eigenvector}\\
\hat{H}|P_{\mathrm{H}}\rangle&=\left(M_{\mathrm{H}}-\frac{i}{2}\Gamma_{\mathrm{H}}\right)|P_\mathrm{H}\rangle,\label{eigenvector2}
\end{align}
where \(M_\mathrm{L},M_\mathrm{H},\Gamma_\mathrm{L},\Gamma_\mathrm{H}\) are real positive numbers specifying the eigenvalues of the two unstable eigenstates, $|P_\mathrm{L}\rangle$ and $|P_\mathrm{H}\rangle$.  Obviously, $M_\mathrm{L}$, $M_\mathrm{H}$ and $\Gamma_\mathrm{L}$, $\Gamma_\mathrm{H}$ are related to the mass and the width of the decay of the respective eigenstates.  
For definiteness, we take $M_\mathrm{L} \leq M_\mathrm{H}$ below.

Because of the non-Hermiticity of the Hamiltonian, these eigenstates $|P_{\rm L}\rangle$, $|P_{\rm H}\rangle$ may not be orthogonal to each other, and one is required to take care of them properly for evaluating the physical values obtained in the measurement.  In particular, it has been argued that, when the measurement is performed through the energy eigenstates, one has to introduce the bi-orthogonal basis defined from the Hamiltonian, resulting in a modification of the weak value \cite{Matzkin_2012}.  This does not apply to our analysis, since our measurement for evaluating the lifetime is assumed to be made through some orthogonal basis states $|P\rangle$, $|\bar{P}\rangle$ at the end of the process.  In computation, this will be done by transcribing the energy eigenstates into these basis states via the relations,
\begin{align}
\ket{P_\mathrm{L}}&=p_{1}|P\rangle+q_{1}|\bar{P}\rangle, \label{plfromb0b0bar}\\
\ket{P_\mathrm{H}}&=p_{2}|P\rangle-q_{2}|\bar{P}\rangle, \label{phfromb0b0bar}
\end{align}
with complex numbers $p_{i}$ and $q_{i}$ satisfying $|p_{i}|^{2}+|q_{i}|^{2}=1$ for $i = 1, 2$.
The choice of the basis states, $|P\rangle$ and $|\bar{P}\rangle$, is immaterial, as it does not affect the possible range of lifetimes realized by various postselections, which is obviously basis-independent.  It may well be convenient, however, to choose the basis states as eigenstates of a Hermitian operator corresponding to a symmetry transformation of the system.  In fact, in the discussion of the $B$ mesons in particle physics, it is customary to choose the orthonormal basis states by the flavor eigenstates. 
Now, let $\hat{D}$ be the operator of interchanging the basis states, 
$\hat{D}|P\rangle = |\bar{P}\rangle$ and $\hat{D}|\bar{P}\rangle = |P\rangle$. 
If $\hat{D}$ happens to be a symmetry of the system, {\it i.e.}, if $[\hat{H}, \hat{D}]=0$, the energy eigenstates are eigenstates of the symmetry transformation $\hat{D}$ simultaneously.  In that case, we may assume $\hat{D}\ket{P_\mathrm{L}} = \pm \ket{P_\mathrm{L}}$ and $\hat{D}\ket{P_\mathrm{H}} = \mp\ket{P_\mathrm{H}}$, which implies
$p_{i}=\pm q_{i} = 1/\sqrt{2}$ up to a common phase.  

Let $|\Psi\rangle$ be the state of the system prepared at the initial time $t = 0$, which is is called preselected state in the context of weak value amplification.  It can be expanded in terms of the energy eigenstates, or similarly of the basis states, as
\begin{align}
|\Psi\rangle 
=a_\mathrm{L}|P_\mathrm{L}\rangle+a_\mathrm{H}|P_\mathrm{H}\rangle 
=a_\mathrm{P}|P\rangle+a_{\bar{\mathrm{P}}}|\bar{P}\rangle,
\label{eq:init-state}
\end{align}
with some coefficients $a_\mathrm{L}$, $a_\mathrm{H}$ and $a_\mathrm{P}$, $a_\mathrm{\bar{P}}$ fulfilling the normalization 
conditions, $|a_\mathrm{P}|^{2}+|a_\mathrm{\bar{P}}|^{2} =1$. 
In view of \eqref{plfromb0b0bar} and \eqref{phfromb0b0bar}, 
they are related by 
\begin{align}
a_\mathrm{P} = a_\mathrm{L}p_{1}+a_\mathrm{H}p_{2},  \quad
a_\mathrm{\bar{P}} = a_\mathrm{L}q_{1}-a_\mathrm{H}q_{2}.
    \label{eq:const3}
\end{align}
At a later time $t$, the state $|\Psi\rangle $ evolves into 
\begin{align}
|\Psi (t)\rangle& =e^{-it\hat{H}}|\Psi\rangle \nonumber\\
&=a_\mathrm{L}e^{-t\frac{\Gamma_\mathrm{L}}{2}-itM_\mathrm{L}}|P_\mathrm{L}\rangle+a_\mathrm{H}e^{-t\frac{\Gamma_\mathrm{H}}{2}-itM_\mathrm{H}}|P_\mathrm{H}\rangle.
\label{evolved-state}
\end{align}
The probability that a particle has not collapsed at time $t$ is then given by the norm,
\begin{align}
    \Vert |\Psi (t)\rangle\Vert^{2}
    &=|a_\mathrm{L}|^{2}e^{-\Gamma_\mathrm{L}t}+|a_\mathrm{H}|^{2}e^{-\Gamma_\mathrm{H}t}\nonumber\\
    &\quad+2e^{-t\Gamma}\mathrm{Re}[a_\mathrm{L}^{\ast}a_\mathrm{H}\braket{P_\mathrm{L}|P_\mathrm{H}}e^{-it\Delta M}],
\end{align}
where we have introduced the mass difference $\Delta M$ and the average width $\Gamma$ 
by 
\begin{align}
\Delta M =M_\mathrm{H}-M_\mathrm{L}, \qquad \Gamma =\frac{\Gamma_\mathrm{L}+\Gamma_\mathrm{H}}{2}.\label{def:deltam_and_gamma}
\end{align}
From these, we obtain the decay time distribution,
\begin{align}
    N(t|\Psi)&=-\frac{d}{dt} \Vert |\Psi (t)\rangle\Vert^{2}\nonumber\\
&=|a_\mathrm{L}|^{2}\Gamma_\mathrm{L}e^{-\Gamma_\mathrm{L}t}+|a_\mathrm{H}|^{2}\Gamma_\mathrm{H}e^{-\Gamma_\mathrm{H}t}\nonumber\\
    &\quad+2\Gamma e^{-t\Gamma}\, \mathrm{Re}\left[a_\mathrm{L}^{\ast}a_\mathrm{H}\braket{P_\mathrm{L}|P_\mathrm{H}}e^{-it\Delta M}\right]\nonumber\\
    &\quad  -2\Delta M e^{-t\Gamma}\, \mathrm{Im}\left[a_\mathrm{L}^{\ast}a_\mathrm{H}\braket{P_\mathrm{L}|P_\mathrm{H}}e^{-it\Delta M}\right],\label{distribution_decay}
\end{align}
which obeys $\int_{0}^{\infty} dt \,N(t|\Psi) = 1$ by construction.

When, in particular, we have the special case where the relations
\begin{align}
    \Gamma_\mathrm{L}=\Gamma_\mathrm{H},\qquad \langle P_\mathrm{H}|P_\mathrm{L}\rangle=0
\end{align}
hold, then the time distribution \eqref{distribution_decay} reduces to
\begin{align}
   N(t|\Psi)=\Gamma e^{-\Gamma t}
       \label{simpledist}
\end{align}
as expected. 

Returning to the general case, we can now evaluate the lifetime $\tau(\Psi)$ of the initial state $\Psi$ 
from the time distribution as 
\begin{align}
    &\tau(\Psi):=\int_{0}^{\infty} dt\ t N(t|\Psi)\nonumber\\
    &=\frac{|a_\mathrm{L}|^{2}}{\Gamma_\mathrm{L}}+\frac{|a_\mathrm{H}|^{2}}{\Gamma_\mathrm{H}}+2\Gamma\, \mathrm{Re}\left[a_\mathrm{L}^{\ast}a_\mathrm{H}\braket{P_\mathrm{L}|P_\mathrm{H}}\right]\frac{\Gamma^{2}-\Delta M^{2}}{(\Gamma^{2}+\Delta M^{2})^{2}}\nonumber\\
    &\quad-2\Delta M\, \mathrm{Im}\left[a_\mathrm{L}^{\ast}a_\mathrm{H}\braket{P_\mathrm{L}|P_\mathrm{H}}\right]\frac{\Gamma^{2}-\Delta M^{2}}{(\Gamma^{2}+\Delta M^{2})^{2}}\nonumber\\
    &\quad+2\Gamma\, \mathrm{Im}\left[a_\mathrm{L}^{\ast}a_\mathrm{H}\braket{P_\mathrm{L}|P_\mathrm{H}}\right]\frac{2\Gamma\Delta M}{(\Gamma^{2}+\Delta M^{2})^{2}}\nonumber\\
    &\quad +2\Delta M\, \mathrm{Re}\left[a_\mathrm{L}^{\ast}a_\mathrm{H}\braket{P_\mathrm{L}|P_\mathrm{H}}\right]\frac{2\Gamma\Delta M}{(\Gamma^{2}+\Delta M^{2})^{2}}.
    \label{def:taup}
\end{align}
Note that, as long as the two eigenstates are non-orthogonal $\braket{P_\mathrm{L}|P_\mathrm{H}} \ne 0$, the lifetime depends on 
the choice of the initial state $\Psi$.

\section{Decay time distribution under Postselection}\label{sec:time_distribution_post}

In this section, we will discuss the time distribution under postselection, but before that, we would like to mention how postselection may be implemented for unstable particles in atomic and particle physics.  
The basic idea of implementing the postselection is to utilize possible correlations between the postselected states and the decay modes.  In atomic physics, this has actually been implemented in an experiment involving atomic decays \cite{Shomroni_2013}, where the polarization of photons emitted in the decay is used to specify the postselected states.  In particle physics, a similar process based on the photon polarization has also been considered for the decay of the $B$ meson \cite{Higashino_2020} with the help of the $CP$ symmetry, where the feasibility of utilizing scattering angles of charged leptons is studied additionally as an alternative.  In the following discussion, we will assume that these schemes allow us to freely choose the postselected state denoted by $|\Phi\rangle$.

Now, we would like to consider how the lifetime will be affected when the postselection of state is made. 
Before the full consideration, let us first see how it could go in a simplified argument.  We first evaluate the transition amplitude at time $t$,
\begin{align}
    \langle \Phi|\Psi (t)\rangle =\langle \Phi|e^{-it\hat{H}}|\Psi\rangle.
    \label{amp_pps}
\end{align}
Upon using the linear approximation in time evolution (whose validity will be discussed shorly), this becomes
\begin{align}
\langle \Phi|\Psi (t)\rangle
    &\simeq \langle \Phi|\left(1-it\hat{H}\right)|\Psi\rangle\nonumber\\
    &=\langle \Phi|\Psi\rangle\left(1-itH_\mathrm{w}\right)\nonumber\\
    &\simeq\langle \Phi|\Psi\rangle e^{t\, \mathrm{Im}[H_\mathrm{w}]-it\, \mathrm{Re}[H_\mathrm{w}]},\label{amp_ll}
\end{align}
where $H_\mathrm{w}$ is the weak value $ H_\mathrm{w} =\langle \Phi|\hat{H}|\Psi\rangle/\langle \Phi|\Psi\rangle$ of the Hamiltonian.
The transition probability is thus obtained as 
\begin{align}
    |\langle \Phi|\Psi (t)\rangle|^{2}&=|\langle \Phi|\Psi\rangle|^{2}e^{2t\, \mathrm{Im}[H_\mathrm{w}]}.
    \label{eq:unit-evolve}
\end{align}

Our argument becomes more transparent by shifting the Hamiltonian
by the amount of the average energy eigenvalue as
\begin{align}
    \hat{H}=\left(M-\frac{i}{2}\Gamma\right)\hat{1}+g\hat{A},
    \label{def:amiltonian}
\end{align}
with
\begin{align}
    M=\frac{M_\mathrm{L}+M_\mathrm{H}}{2}, \qquad g=\frac{\Delta M}{2}
    \label{average-mass}
\end{align}
and thereby introduce the \lq normalized\rq\ Hamiltonian,
\begin{align}
\hat{A} = \frac{1}{g}\left[ \hat{H} - \left( M-\frac{i}{2}\Gamma\right)\hat{1}\right].
\label{eq:deltah}
\end{align}
As we can see in \eqref{def:amiltonian}, the normalized Hamiltonian $\hat{A}$ describes the rescaled time evolution relative to the 
average (uniform) evolution characterized by $M$ and $i\Gamma/2$ in the unit of $g$ which renders $\hat{A}$ dimensionless.  The constant $g$ provides the scale of the evolution prescribed by $\hat{H}$
and also plays the role of the coupling constant in the usual discussion of weak value amplification. 
It is notable that the operator $\hat{A}$, whose eigenvalues become precisely $+1$ and $-1$ when $\Gamma_\mathrm{L}=\Gamma_\mathrm{H}$, 
resembles the familiar Pauli matrix $\sigma_z$. 

With this preparation, we find that the transition amplitude \eqref{amp_pps} becomes
\begin{align}
    \langle \Phi|\Psi (t)\rangle&=e^{-t\frac{\Gamma}{2}}e^{-itM}\langle \Phi|e^{-itg\hat{A}}|\Psi\rangle.
    \label{eq:transres}
\end{align}
Observe that 
the linear approximation we used from \eqref{amp_pps} to \eqref{amp_ll}, which now applies to \eqref{eq:transres}, is valid as long as
\begin{align}
tg = t\Gamma \left(\frac{g}{\Gamma}\right) = t\Gamma \left(\frac{\Delta M}{2\Gamma}\right) \ll 1
\label{eq:cvalcon}
\end{align}
is valid.
Notice that, on account of the overall exponential factor $e^{-t\frac{\Gamma}{2}}$ in
\eqref{eq:transres} which (when squared) suppresses the probability for $t > 1/\Gamma$, the range of time $t$ we need to consider is practically confined to 
$t\Gamma \le 1$.  This implies that the condition \eqref{eq:cvalcon} holds if $\Delta M/\Gamma \ll 1$ or 
\begin{align}
\Delta M \ll \Gamma.
\label{eq:appr}
\end{align}
Since $g$ acts as the coupling constant for time evolution as we noted earlier, the condition \eqref{eq:appr} which ensures \eqref{eq:cvalcon} 
simply means that we are working in the slow range of evolution, which corresponds to 
the weak limit of physical interaction in the usual context of weak value amplification.  

When this condition holds, the resulting transition probability is given, instead of \eqref{eq:unit-evolve}, by
\begin{align}
    |\langle \Phi|\Psi (t)\rangle|^{2}&=e^{-t\Gamma}|\langle \Phi|\Psi\rangle|^{2}e^{2tg\mathrm{Im}[A_\mathrm{w}]},
    \label{eq:unit-evolrev}
\end{align}
where 
\begin{align}
A_\mathrm{w} = \frac{\langle \Phi|\hat{A}|\Psi\rangle}{\langle \Phi|\Psi\rangle} 
\label{eq:awv}
\end{align}
is the weak value of the normalized Hamiltonian \eqref{eq:deltah}.
Clearly, the result \eqref{eq:unit-evolrev} remains to be valid even for general cases with more than two energy eigenstates as long as the 
Hamiltonian is bounded, once we implement the normalization \eqref{eq:deltah} with $M$ being the average of the entire eigenvalues.  

In order to find the distribution of decay time when the postselection is made for general cases, not just for the case \eqref{eq:appr}, 
as we did for the preselected state in \eqref{eq:init-state} we first expand the postselected state $|\Phi\rangle$ as 
\begin{align}
|\Phi\rangle =b_\mathrm{L}|P_\mathrm{L}\rangle+b_\mathrm{H}|P_\mathrm{H}\rangle
=b_\mathrm{P}|P\rangle+b_\mathrm{\bar{P}}|\bar{P}\rangle,
\label{eq:post-state}
\end{align}
with coefficients, $b_\mathrm{L}$, $b_\mathrm{H}$, $b_\mathrm{P}$ and $b_\mathrm{\bar{P}}$ fulfilling the normalization conditions, 
$|b_\mathrm{P}|^{2}+|b_\mathrm{\bar{P}}|^{2}=1$.
From \eqref{plfromb0b0bar} and \eqref{phfromb0b0bar}, 
they are related by $b_\mathrm{P}=b_\mathrm{L}p_{1}+b_\mathrm{H}p_{2}$ and $b_\mathrm{\bar{P}}=b_\mathrm{L}q_{1}-b_\mathrm{H}q_{2}$.
Then the probability amplitude that the time evolved state $|\Psi (t)\rangle$ can be found in the postselected state is
\begin{align}
    \langle \Phi|\Psi (t)\rangle&=\left(b_\mathrm{L}^{\ast}+b_\mathrm{H}^{\ast}\langle P_\mathrm{H}|P_\mathrm{L}\rangle\right)a_\mathrm{L}e^{-t\frac{\Gamma_\mathrm{L}}{2}-itM_\mathrm{L}}\nonumber\\
    &\quad+\left(b_\mathrm{H}^{\ast}+b_\mathrm{L}^{\ast}\langle P_\mathrm{L}|P_\mathrm{H}\rangle \right)a_\mathrm{H}e^{-t\frac{\Gamma_\mathrm{H}}{2}-itM_\mathrm{H}}.\label{eq:inpro}
\end{align}
With the shorthand,
\begin{align}
    c_\mathrm{L} &=\left(b_\mathrm{L}^{\ast}+b_\mathrm{H}^{\ast}\langle P_\mathrm{H}|P_\mathrm{L}\rangle\right)a_\mathrm{L}, \label{eq:sha}\\
    c_\mathrm{H} &=\left(b_\mathrm{H}^{\ast}+b_\mathrm{L}^{\ast}\langle P_\mathrm{L}|P_\mathrm{H}\rangle \right)a_\mathrm{H}, \label{eq:shb}
\end{align}
the corresponding probability becomes
\begin{align}
     |\langle \Phi|\Psi (t)\rangle|^{2}&=|c_\mathrm{L}|^{2}e^{-\Gamma_\mathrm{L}t}+|c_\mathrm{H}|^{2}e^{-\Gamma_\mathrm{H}t}\nonumber\\
     &\quad+2e^{-\Gamma t}\, \mathrm{Re}\left[c_\mathrm{L}^{\ast}c_\mathrm{H}e^{-it\Delta M}\right].
     \label{eq:dm}
\end{align}
The decay time distribution conditioned by the postselection may then be defined by
\begin{align}
N(t|\Psi\to \Phi) = \frac{|\langle \Phi|\Psi (t)\rangle|^{2}}{\int_{0}^{\infty} dt\ |\langle \Phi|\Psi (t)\rangle|^{2}},
\label{eq:timedistcond}
\end{align}
from which the conditional lifetime $\tau(\Psi\to \Phi)$ follows as
\begin{align}
   \tau(\Psi\to \Phi)&:= \int_{0}^{\infty} dt\ t N(t|\Psi\to \Phi) \nonumber\\
   &\,\, =\frac{\int_{0}^{\infty} dt\ t |\langle \Phi|\Psi (t)\rangle|^{2}}{\int_{0}^{\infty} dt\ |\langle \Phi|\Psi (t)\rangle|^{2}}.
   \label{eq:cond-lifetime}
\end{align}

Before proceeding further, we present here the closed form of the weak value \eqref{eq:awv}
evaluated with the help of \eqref{eq:init-state}, \eqref{eq:post-state}, \eqref{eq:sha} and \eqref{eq:shb},
\begin{align}
    A_\mathrm{w} 
     =\frac{c_\mathrm{H}-c_\mathrm{L}}{c_\mathrm{H}+c_\mathrm{L}}\left(1-\frac{i}{2}\frac{\Gamma_{\mathrm{H}}-\Gamma_\mathrm{L}}{\Delta M}\right)
    \label{weak_val_dmdg}
\end{align}
whose real and imaginary parts are, respectively,
\begin{align}
    \mathrm{Re}\left[A_\mathrm{ w}\right]
    &=\frac{|c_\mathrm{H}|^{2}-|c_\mathrm{L}|^{2}}{|c_\mathrm{H}+c_\mathrm{L}|^{2}}+\frac{\mathrm{Im}[c_\mathrm{L}^{\ast}c_\mathrm{H}]}{|c_\mathrm{H}+c_\mathrm{L}|^{2}}\frac{\Gamma_\mathrm{H}-\Gamma_\mathrm{L}}{\Delta M}, \label{realpart_weakvalue}\\
    \mathrm{Im}\left[A_\mathrm{w}\right]
    &=-\frac{|c_\mathrm{H}|^{2}-|c_\mathrm{L}|^{2}}{|c_\mathrm{H}+c_\mathrm{L}|^{2}}\frac{\Gamma_\mathrm{H}-\Gamma_\mathrm{L}}{2\Delta M}+2\frac{\mathrm{Im}[c_\mathrm{L}^{\ast}c_\mathrm{H}]}{|c_\mathrm{H}+c_\mathrm{L}|^{2}}.
    \label{imaginarypart_weakvalue}
\end{align}

In the usual discussion of weak value amplification, the amplification effect is realized by choosing the postselection such that 
the weak value exceeds the range of the eigenvalues of the physical observable that one wishes to amplify in the measurement.  
In the present case, the observable corresponds to the (normalized) Hamiltonian \eqref{eq:deltah}, and the discussion on the extent of
amplification requires the consideration of the particular form of the observable as well as the 
postselection we shall make.  This will be our topic of the next section.

\section{Lifetime and its range}\label{sec:lifetime_and_its_range}

As can be seen from \eqref{eq:unit-evolve} which is valid in the linear approximation, the transition probability is directly affected by (the imaginary part of) the weak value of the Hamiltonian.   The lifetime \eqref{eq:cond-lifetime} is then found to be
\begin{align}
\tau(\Psi\to \Phi) \simeq \frac{1}{\Gamma} +\frac{2g}{\Gamma^2}\,\mathrm{Im}[A_\mathrm{w}],
\label{effective_Gamma}
\end{align}
which shows that the lifetime of a decaying particle can effectively be altered by varying the postselected state $\ket{\Phi}$ on which the weak value $A_\mathrm{w}$ depends.   This should be reasonable because $A_\mathrm{w}$ is the weak value of the normalized Hamiltonian $\hat A$ which dictates the time evolution of the system \eqref{eq:transres}.  In the context of weak value amplification applied to the present system, one may say that the lifetime may be enlarged by tuning the postselection properly relative to the constant $g$.  

Now, the possible range of the lifetime realized by the variation of postselection under the given preselected state $\ket{\Psi}$ can be found by evaluating the conditional lifetime \eqref{eq:cond-lifetime} without resorting to the linear approximation.  In the two dimensional case we are considering, this can be done easily (see Appendix A) and the result is
\begin{widetext}
\begin{align}
    \tau(\Psi\to \Phi)=\frac{\frac{|c_\mathrm{L}|^{2}}{\Gamma^{2}_\mathrm{L}}+\frac{|c_\mathrm{H}|^{2}}{\Gamma^{2}_\mathrm{H}} +\frac{2(\Gamma^2-\Delta M^{2})}{(\Gamma^{2}+\Delta M^{2})^{2}}\mathrm{Re}\left[c_\mathrm{L}^{\ast}c_\mathrm{H}\right]+\frac{4\Gamma\Delta M}{(\Gamma^{2}+\Delta M^{2})^{2}}\mathrm{Im}\left[c_\mathrm{L}^{\ast}c_\mathrm{H}\right]}{\frac{|c_\mathrm{L}|^{2}}{\Gamma_\mathrm{L}}+\frac{|c_\mathrm{H}|^{2}}{\Gamma_\mathrm{H}}+\frac{2\Gamma}{\Gamma^{2}+\Delta M^{2}}\mathrm{Re}\left[c_\mathrm{L}^{\ast}c_\mathrm{H}\right]+\frac{2\Delta M}{\Gamma^{2}+\Delta M^{2}}\mathrm{Im}\left[c_\mathrm{L}^{\ast}c_\mathrm{H}\right]}.
\label{act_lifetime}
\end{align}
\end{widetext}

To avoid unnecessary complexity and yet keep the essential ingredient of the matter, let us assume that $p_{1}=p_{2}=p$ and $q_{1}=q_{2}=q$.  
As we mentioned before,  this is satisfied when, for instance, we have the symmetry under the interchange $\hat D$, and one prominent example of this is the system of decaying $B$ mesons, where the role of the $\hat D$ symmetry is played by the $CPT$ symmetry.  
Under this assumption, we have
\begin{align}
    \braket{P_\mathrm{H}|P_\mathrm{L}}=|p|^{2}-|q|^{2}.
\end{align}

To proceed further, we introduce the shorthand,
\begin{align}
    k=\frac{\Gamma_\mathrm{H}}{\Gamma_\mathrm{L}},
\end{align}
and the ratio,
\begin{align}
    R=\frac{\tau(\Psi\to\Phi)}{\tau(\Psi)},
    \label{def:R}
\end{align}
which signifies the rate of variation the lifetime has acquired by the postselection compared to the case where no postselection is made.  Note that, for considering the effect of postselection, one has to express $R$ as a function of 
$(p, q, b_\mathrm{P},b_\mathrm{\bar{P}},k,\Gamma,\Delta M)$, where the dependence on $b_\mathrm{P}$ and $b_\mathrm{\bar{P}}$ are given by $c_\mathrm{L}$ and $c_\mathrm{H}$ in \eqref{eq:sha} and \eqref{eq:shb} (see Appendix A for the formula to be used).

For convenience, 
we define the basis states $|P\rangle$ and $|\bar{P}\rangle$ such that both $p$ and $q$ become real, which is always possible by absorbing the phase factor into the respective states.  
Once this is done, then with respect to the basis states we may define the relative phase $\theta$ between $b_\mathrm{P}$ and $b_\mathrm{\bar{P}}$ by
\begin{align}
    b_\mathrm{P}b_\mathrm{\bar{P}}^{*}=|b_\mathrm{P}||b_\mathrm{\bar{P}}|e^{i\theta}.
\end{align}
Thus, taking account of the constraints coming from the normalization conditions, we recognize that  
the ratio $R$ can now be regarded as a function of  
$(p,|b_\mathrm{P}|,\theta, k, \Gamma, \Delta M)$. 

To solve the constraints, in the following discussion we restrict ourselves to the particular case,
\begin{align}
    p=q=\frac{1}{\sqrt{2}},
    \label{restrict}
\end{align}
implying that the two eigenstates are orthogonal to each other.  We also consider, for definiteness, the case where $a_\mathrm{P}=1$ and $a_{\bar{\mathrm{P}}}=0$, {\it i.e.}, the case where the initial state is given by one of the eigenstates as
\begin{align}
|\Psi\rangle 
=|P\rangle
=\frac{1}{\sqrt{2}}\left(|P_\mathrm{H}\rangle+|P_\mathrm{L}\rangle\right). \label{restrictpre}
\end{align}
One still expects that this restriction will not lose track of the basic feature of lifetime we are studying, partly because the initial state $|\Psi\rangle$ is now a maximal superposition of the two eigenstates, and partly 
because the lifetime is determined essentially with respect to the postselected state $|\Phi\rangle$ which are to be varied with the parameters introduced earlier.  

With \eqref{restrict} and \eqref{restrictpre} we obtain 
\begin{align}
\tau(\Psi)=\frac{1}{2\Gamma_\mathrm{L}}+\frac{1}{2\Gamma_\mathrm{H}}
=\frac{(1+k)^{2}}{4k\Gamma}
\label{eq:tau}
\end{align}
and also
\begin{widetext}
\begin{align}
   &\tau(\Psi\to\Phi) \nonumber\\
&=\frac{(\frac{1}{4}+\frac{1}{2}|b_\mathrm{P}||b_\mathrm{\bar{P}}|\cos\theta)\frac{(1+k)^{2}}{4\Gamma^{2}}+(\frac{1}{4}-\frac{1}{2}|b_\mathrm{P}||b_\mathrm{\bar{P}}|\cos\theta)\frac{(1+k)^{2}}{4k^{2}\Gamma^{2}}+\frac{\Gamma^{2}-\Delta M^{2}}{(\Gamma^{2}+\Delta M^{2})^{2}}\cdot\frac{|b_\mathrm{P}|^{2}-|b_\mathrm{\bar{P}}|^{2}}{2}-\frac{2\Gamma\Delta M}{(\Gamma^{2}+\Delta M^{2})^{2}}|b_\mathrm{P}||b_\mathrm{\bar{P}}|\sin\theta}{(\frac{1}{4}+\frac{1}{2}|b_\mathrm{P}||b_\mathrm{\bar{P}}|\cos\theta)\frac{1+k}{2\Gamma}+(\frac{1}{4}-\frac{1}{2}|b_\mathrm{P}||b_\mathrm{\bar{P}}|\cos\theta)\frac{1+k}{2k\Gamma}+\frac{\Gamma}{\Gamma^{2}+\Delta M^{2}}\cdot\frac{|b_\mathrm{P}|^{2}-|b_\mathrm{\bar{P}}|^{2}}{2}-\frac{\Delta M}{\Gamma^{2}+\Delta M^{2}}|b_\mathrm{P}||b_\mathrm{\bar{P}}|\sin\theta},\label{eff-lifetime}
\end{align}
from which the ratio $R$ is found to be
\begin{align}
  R
  &=\frac{(\frac{1}{4}+\frac{1}{2}|b_\mathrm{P}||b_\mathrm{\bar{P}}|\cos\theta)k+(\frac{1}{4}-\frac{1}{2}|b_\mathrm{P}||b_\mathrm{\bar{P}}|\cos\theta)\frac{1}{k}+2k\Gamma^{2}\frac{\Gamma^{2}-\Delta M^{2}}{(\Gamma^{2}+\Delta M^{2})^{2}}\cdot\frac{|b_\mathrm{P}|^{2}-|b_\mathrm{\bar{P}}|^{2}}{(1+k)^{2}}-\frac{8k\Gamma^{3}\Delta M}{(1+k)^{2}(\Gamma^{2}+\Delta M^{2})^{2}}|b_\mathrm{P}||b_\mathrm{\bar{P}}|\sin\theta}{(\frac{1}{4}+\frac{1}{2}|b_\mathrm{P}||b_\mathrm{\bar{P}}|\cos\theta)\frac{1+k}{2}+(\frac{1}{4}-\frac{1}{2}|b_\mathrm{P}||b_\mathrm{\bar{P}}|\cos\theta)\frac{1+k}{2k}+\frac{\Gamma^{2}}{\Gamma^{2}+\Delta M^{2}}\cdot\frac{|b_\mathrm{P}|^{2}-|b_\mathrm{\bar{P}}|^{2}}{2}-\frac{\Gamma\Delta M}{\Gamma^{2}+\Delta M^{2}}|b_\mathrm{P}||b_\mathrm{\bar{P}}|\sin\theta}.
\label{fullratio}
\end{align}
\end{widetext}
At this point, we recognize that both \eqref{eff-lifetime} and \eqref{fullratio} are invariant under the simultaneous transformation $k \to 1/k$ and $\theta \to \pi - \theta$ as a consequence of the invariance under the formal exchange of two eigenstates.  This implies that, when we investigate the behavior of $\tau(\Psi\to\Phi)$ or $R$, it is sufficient to consider the range $1 \le k$ since the range $0< k< 1$ can be obtained from the former by the above transformation.  We shall therefore discuss only the range $1 \le k$ in our analysis below.  
Note that if we consider the extreme case $k \to \infty$ while keeping $\Gamma$ constant, where we have  $\Gamma_\mathrm{L} \to 0$ and $\Gamma_\mathrm{H} \to 2\Gamma$, the lifetime \eqref{eq:tau} tends to diverge $\tau(\Psi) \to \infty$.  This is, of course, a consequence of the fact that the component $|P_\mathrm{L}\rangle$ in the state \eqref{restrictpre} ceases to decay in the limit.  

We now examine how much we can vary the ratio $R$ by choosing different 
parameters $b_\mathrm{P}$ and $\theta$ for the postselection. 
Prior to this, let us consider the special case,
\begin{align} 
|b_\mathrm{P}|=|b_\mathrm{\bar{P}}|=\frac{1}{\sqrt{2}}, \quad \theta=\pi,
    \label{sppsco}
\end{align}
which amounts to the postselection,
\begin{align}
|\Phi\rangle 
=\frac{1}{\sqrt{2}}\left(|P\rangle-|\bar{P}\rangle\right)
=|P_\mathrm{H}\rangle,
    \label{sppostsel}
\end{align}
up to an overall phase. 
We then find that the lifetime under this postselection becomes 
\begin{align}
\tau(\Psi\to\Phi) = \frac{1}{\Gamma_\mathrm{H}} = \frac{1+k}{2k\Gamma},
\label{eq:tautwo}
\end{align}
and hence from \eqref{eq:tau} the ratio $R$ turns out to be
\begin{align}
    R=\frac{2}{1+k}=\frac{\Gamma_\mathrm{L}}{\Gamma}.
    \label{clmin}
\end{align}
In this special case, we find that, at $k = 1$ we have $R = 1$, that is, no net change from the original lifetime by taking the postselection.  
We also observe that,
for $k \to \infty$ we have $R \to 0$, which occurs because our postselection \eqref{sppostsel} filters out the dominant long-lived component $|P_\mathrm{L}\rangle$ completely.  Otherwise, for generic values of the postselection parameters $|b_\mathrm{P}|$ and $\theta$, one can readily confirm the (almost) universal limit $R \to 2$ for $k \to\infty$ as we shall see shortly in the various examples. 

For our later convenience, we also record here that,  
with the parameters $|b_\mathrm{P}|$,  $\theta$
and $k$, the weak value of the normalized Hamiltonian admits the simple form,
\begin{align}
    \mathrm{Re}[A_\mathrm{w}]&=-\frac{4}{|b_\mathrm{P}|^{2}}\cos\theta-\frac{(k-1)\Gamma}{(k+1)\Delta M}\frac{\sqrt{1-|b_\mathrm{P}|^{2}}}{|b_\mathrm{P}|}\sin\theta,  \label{wvre}\\
    \mathrm{Im}[A_\mathrm{w}]&=-\frac{2}{|b_\mathrm{P}|}\sqrt{1-|b_\mathrm{P}|^{2}}\sin\theta.
    \label{wvim}
\end{align}
In what follows, we shall consider three extreme cases for relative values of $\Delta M$ with respect to $\Gamma$, namely, the cases $\Delta M\ll \Gamma$, $\Delta M\simeq\Gamma$ and $\Delta M\gg\Gamma$ in order to examine explicitly the range of lifetimes realized under the variation of the postselection.   As we mentioned in Sec. III,  the $D$ meson system is a concrete example of this case.

\subsection{$\Delta M\ll \Gamma$ case}

Returning to the general case of postselection \eqref{eq:post-state}, we shall first consider 
the case where the mass difference $\Delta M$ is negligible compared to the average width $\Gamma$.  This is actually the case we mentioned in Sec.~III when we employed the linear approximation. 

Specializing to this case,  in the limit $\Delta M/\Gamma \to 0$ we find that the ratio $R$ in \eqref{fullratio} tends to the value,
\begin{widetext}
\begin{align}
    R=\frac{\frac{(k-1)^{2}(k^{2}+4k+1)}{4(1+k)^{2}}+\frac{k^2-1}{2}|b_\mathrm{P}|\sqrt{1-|b_\mathrm{P}|^{2}|}\cos\theta+\frac{4k^2|b_\mathrm{P}|^{2}}{(1+k)^{2}}}{\frac{1}{8}(k-1)^{2}+\frac{k^2-1}{4}|b_\mathrm{P}|\sqrt{1-|b_\mathrm{P}|^{2}}\cos\theta+k|b_\mathrm{P}|^{2}},
    \label{rint}
\end{align}
\end{widetext}
where we immediately confirm $R=1$ at $k=1$ except for the case $|b_\mathrm{P}|=0$. 

\begin{figure}[h]
    \centering
    \includegraphics[width=8.6cm]{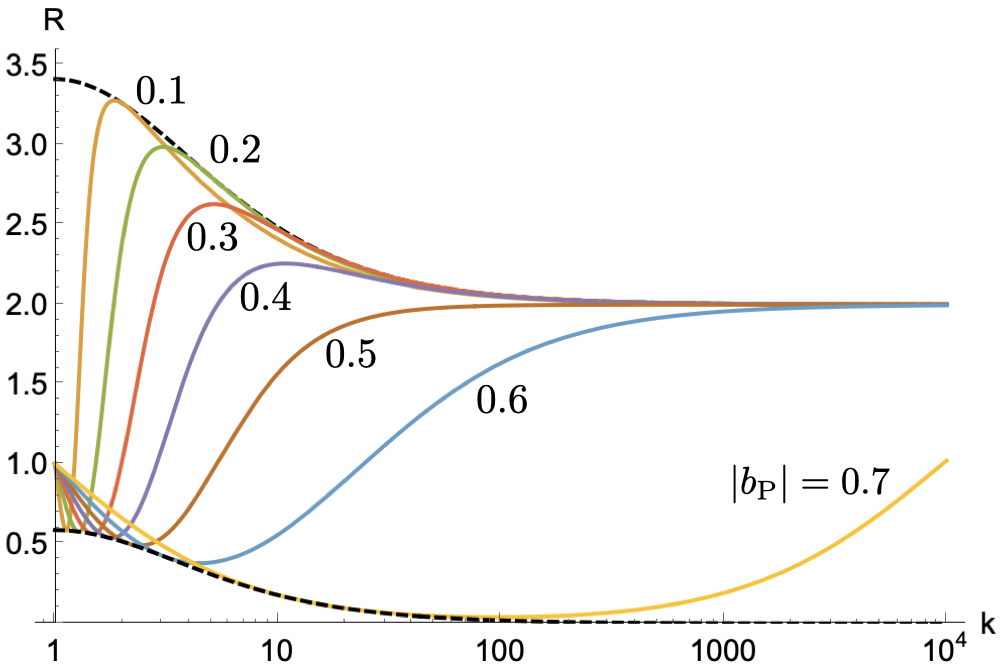}
    \caption{The ratio $R$ as functions of $k$ for seven values of $|b_\mathrm{P}|$ ranging from $0.1$ to $0.7$ for every step $0.1$ under the values $\Delta M/\Gamma=1/1000$ and $\theta=\pi$.  Their upper and lower enveloping curves are given by the dashed lines.}
    \label{fig:case_a}
\end{figure}

The ratio $R$ in \eqref{rint} varies according to the postselection we choose, and the overall trend of its variation may be 
learned from the $R-k$ curves in FIG.\ref{fig:case_a} which correspond to different choices of $|b_\mathrm{P}|$ and $\theta$.  
The upper and lower envelopes of the curves indicate that 
the largest value of $R$ is realized when $k$ is close to $1$, even though putting $k=1$ yields $R=1$ generically.

We observe, however, that the limit $k \to 1$ is quite intriguing in the following respects:  as long as $k \ne 1$, the ratio $R$ approaches $1$ when $|b_\mathrm{P}|$ is made closer to $0$, but in the simultaneous limit of $k \to 1$ and $|b_\mathrm{P}|\to 0$ while the ratio,
 \begin{align}
     x=\frac{|b_\mathrm{P}|}{k-1},
 \end{align}
being kept finite, we have
 \begin{align}
    R=\frac{\frac{3}{8}+\cos\theta \, x+x^{2}}{\frac{1}{8}+\frac{1}{2}\cos\theta \, x+x^{2}},
    \label{fig:rlimit}   
\end{align}
which can be different from the value $1$ we just argued.  In fact, 
we find that the ratio $R$ in \eqref{fig:rlimit} attains the maximal value,
\begin{align}
    R_\mathrm{max}=2+\sqrt{2}\simeq 3.414,
\end{align}
at
\begin{align}
    x=\frac{2-\sqrt{2}}{4},\quad \theta=\pi,
\end{align}
and also the minimal value,
\begin{align}
    R_\mathrm{min}=2-\sqrt{2}\simeq 0.5858, 
\end{align}
at
\begin{align}
    x=\frac{2+\sqrt{2}}{4}, \quad \theta=\pi.
\end{align}
These values constitute, respectively, the upper and lower points on the envelope curves borders at $k = 1$ in FIG.\ref{fig:case_a}.   
Although these values may be difficult to attain as they require specification of the postselected state with exact accuracy, 
they give us the possible range of the ratio concerning the lifetime variations by their limiting values.  
These minimal and maximal values are achieved when the weak value of the Hamiltonian becomes divergent, as can be seen immediately by taking the $|b_\mathrm{P}|\to 0$ limit in \eqref{wvre} and \eqref{wvim}. 

On the other hand, we also observe in FIG.\ref{fig:case_a}
that $R \to 2$ in the limit $k \to \infty$ but the speed of the approach becomes significantly slower as $|b_\mathrm{P}|$ tends to $1/\sqrt{2} \simeq 0.71$.  This is due to the singular effect that occurs precisely at $|b_\mathrm{P}| = 1/\sqrt{2}$ with $\theta = \pi$ as we mentioned before.

An example of this case is furnished by the $D$ meson, where we have $\Delta M/\Gamma\sim 10^{-3}$ and $k$ being close to $1$ according to the review \cite{PDG_2020}. 

\subsection{$\Delta M\simeq\Gamma$ case}

\begin{figure}[htbp]
    \centering
    \includegraphics[width=8.6cm]{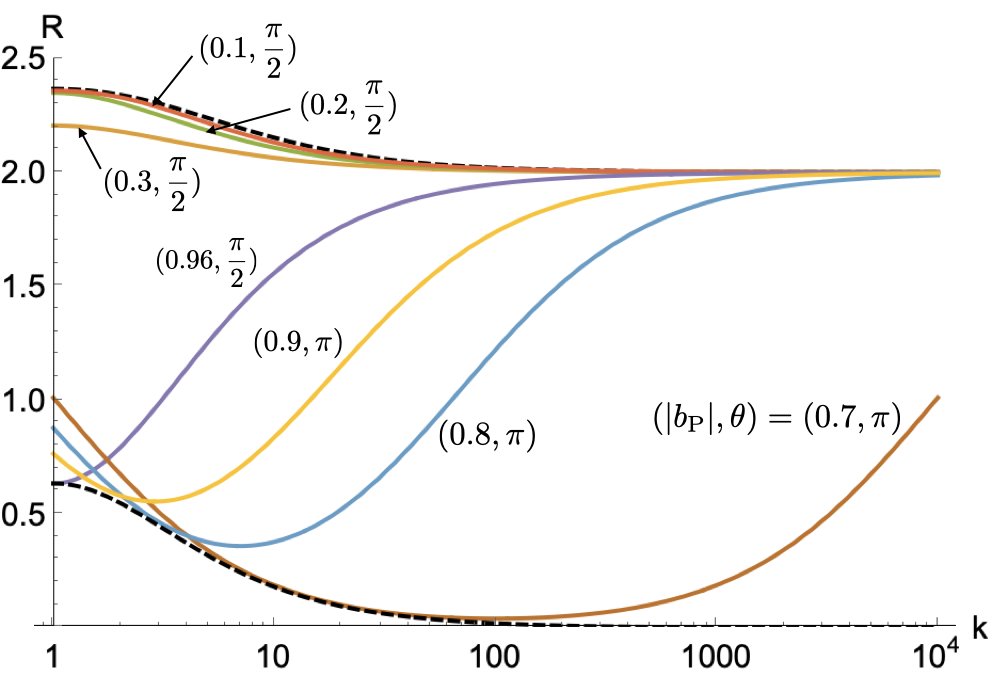}
    \caption{The ratio $R$ as functions of $k$ for various combinations of $(|b_\mathrm{P}|,\theta)$.  Here, we set $\Delta M/\Gamma=1/1.01$ and used $|b_\mathrm{P}|=0.1$, $0.2$, $0.3$, and $0.96$ (giving a value very close to the minimum of $R$ at $k=1$) for $\theta=\pi/2$, and $|b_\mathrm{P}|=0.7$, $0.8$, and $0.9$ for $\theta=\pi$.  Their upper and lower enveloping curves are given by the dashed lines.}
    \label{fig:caseb}
\end{figure}

Next, we consider the case where $\Delta M$ is comparable to $\Gamma$. 
To analyze this, we simply put
$\Delta M = \Gamma$ in \eqref{fullratio} to obtain
\begin{widetext}
\begin{align}
    R=\frac{2\left[(1+k)^{2}(1+k^{2})+2|b_\mathrm{P}|\sqrt{1-|b_\mathrm{P}|^{2}}(k-1)(1+k)^{3}\cos\theta-8k^{2}|b_\mathrm{P}|\sqrt{1-|b_\mathrm{P}|^{2}}\sin\theta)\right]}{(1+k^{2})\left[1+k^{2}+4k|b_\mathrm{P}|^{2}+2|b_\mathrm{P}|\sqrt{1-|b_\mathrm{P}|^{2}}(k^{2}-1)\cos\theta-4k|b_\mathrm{P}|\sqrt{1-|b_\mathrm{P}|^{2}}\sin\theta\right]}.
\end{align}
\end{widetext}
We find that, like in the previous case, in the limit $k \to 1$
the ratio attains its maximal value,
\begin{align}
    R_\mathrm{max}=\frac{3+\sqrt{3}}{2}\simeq 2.366,
\end{align}
at
\begin{align}
   |b_\mathrm{P}|=\frac{\sqrt{2-\sqrt{3}}}{2},\quad \theta=\frac{\pi}{2},
\end{align}
and the minimal value,
\begin{align}
    R_\mathrm{min}=\frac{3-\sqrt{3}}{2}\simeq 0.6340,
\end{align}
at
\begin{align}
   |b_\mathrm{P}|=\frac{\sqrt{2+\sqrt{3}}}{2},\quad \theta=\frac{\pi}{2}.
\end{align}
The $R-k$ curves will then look like FIG. \ref{fig:caseb}, and depending on the choice of $|b_\mathrm{P}|$ and $\theta$, the value at $k=1$ may fall below $1$ or exceed above $2$, while it converges to $2$ in the limit of large $k$ as in the case of $\Delta M\ll\Gamma$.

Let us check the weak value of the Hamiltonian. When the ratio $R$ is minimized, we have
\begin{align}
   \mathrm{Re}[A_\mathrm{w}]&=0,\\
   \mathrm{Im}[A_\mathrm{w}]&=-(4-2\sqrt{3}).
\end{align}
Also, when the ratio is maximized, we have
\begin{align}
    \mathrm{Re}[A_\mathrm{w}]&=0,\\
   \mathrm{Im}[A_\mathrm{w}]&=-(4+2\sqrt{3}).
\end{align}
Unlike the previous case, the ratio $R$ is not optimized when the weak value (especially its imaginary part) tends to diverge. 
This is reminiscent of the behavior we have witnessed when the optimization occurs at finite coupling constants in the weak value amplification experiments \cite{koiketanaka_11, Mori_2019}.

We mention that the $B$ meson is an example of this case, for which $\Delta M/\Gamma \simeq 0.77$ and $k  \simeq 1$. 
In this case, the maximum of $R$ is $2.64$ and the minimum is $0.713$ (for a detailed analysis, see \cite{Higashino_2020}).

\subsection{$\Delta M\gg\Gamma$ case}

Finally, we consider the opposite extreme case, $\Delta M\gg\Gamma$.  In this limit, the ratio $R$ in \eqref{fullratio} admits the simple form,
\begin{align}
R
=\frac{\frac{1}{4}(k^{2}+1)+\frac{1}{2}(k^{2}-1)|b_\mathrm{P}|\sqrt{1-|b_\mathrm{P}|^{2}}\cos\theta}{\frac{1}{8}(1+k)^{2}+\frac{1}{4}(k^{2}-1)|b_\mathrm{P}|\sqrt{1-|b_\mathrm{P}|^{2}}\cos\theta},\label{r:casec}
\end{align}
One then finds that the ratio $R$ takes the maximal value,
\begin{align}
    R_\mathrm{max}=\frac{2k}{1+k}=\frac{\Gamma_\mathrm{H}}{\Gamma},
\end{align}
at 
\begin{align}
    |b_\mathrm{P}|=\frac{1}{\sqrt{2}},\quad \theta=0,
\end{align}
and the minimal value,
\begin{align}
    R_\mathrm{min}=\frac{2}{1+k}=\frac{\Gamma_\mathrm{L}}{\Gamma},
\end{align}
at the specific case \eqref{sppsco}.

\begin{figure}[htbp]
    \centering
    \includegraphics[width=8.6cm]{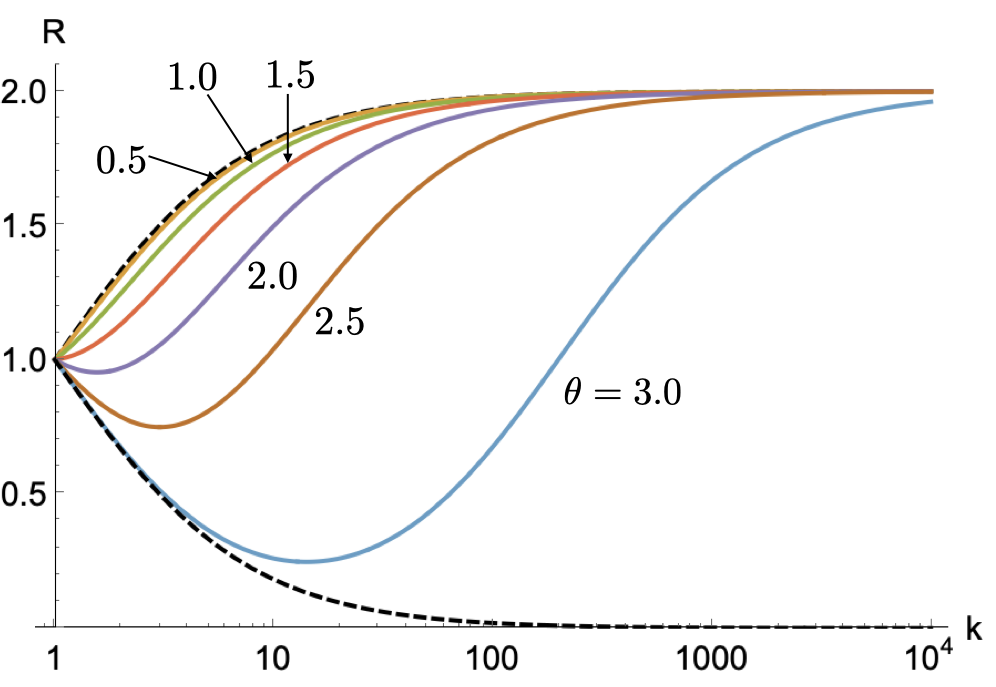}
    \caption{The ratio $R$ as functions of $k$ for various $\theta$ from $0.0$ to $3.0$ in increments of $0.5$ under the values $\Delta M/\Gamma = 1000$ and $|b_\mathrm{P}| = 1/\sqrt{2}$. Their upper and lower enveloping curves are given by the dashed lines.}
    \label{fig:casec}
\end{figure}

As can be seen by substituting $1$ for $k$ in \eqref{r:casec}, $R = 1$ universally for $k = 1$.  Also, as mentioned earlier, in the limit $k \to \infty$, we have $R \to 2$ except for the singular case \eqref{sppsco}. Indeed, in FIG. \ref{fig:casec}, we observe that when $\theta$ is chosen to be $3.0$ (which is close to $\pi$), $R$ begins to grow slowly but will eventually converge to $2$.  

This result is, in fact, identical to that obtained when the oscillation between the long-lived and short-lived states of the particle is negligible and hence only a classical probabilistic mixing of the two will be observed.  This can also be confirmed by examining
the classical limit by restoring the Planck constant $\hbar$ explicitly and rewrite $\Delta M \Rightarrow \Delta M/\hbar$.  This amounts to
\begin{align}
    \frac{\Delta M}{\Gamma}\Rightarrow\frac{\Delta M}{\hbar\Gamma},
\end{align}
which indicates that the classical limit $\hbar \to 0$ is, in effect, equivalent to the limit $\Delta M/\Gamma \to \infty$, which is no other than the present case we are considering. 

The B meson containing a strange quark, {\it i.e.}, the $B^{0}_{S}$ meson, exemplifies the present case, where we have $\Delta M/\Gamma=26.89\pm0.07$ \cite{PDG_2020}.

\subsection{Overall picture of lifetime amplification}

\begin{figure}[htbp]
\centering
\includegraphics[width=8.6cm]{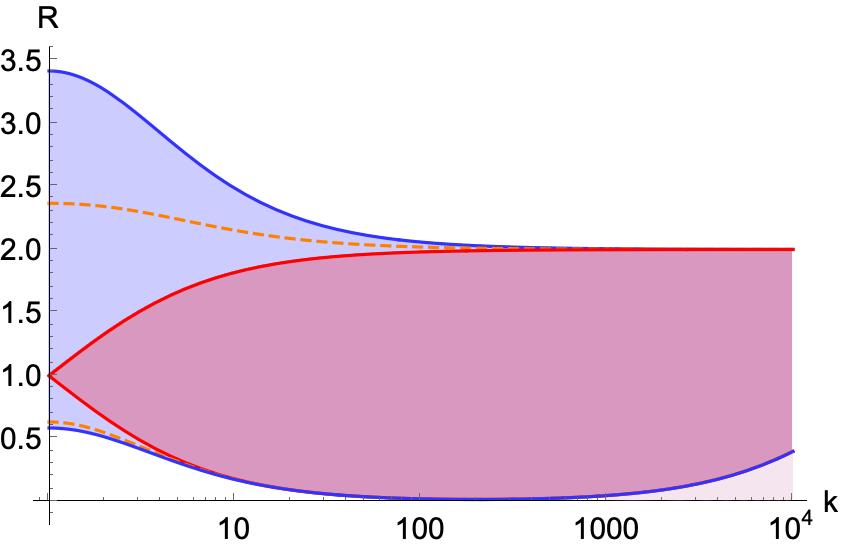}
\caption{Possible regions of the ratio $R$ for the three cases $\Delta M\ll\Gamma$, $\Delta M\gg\Gamma$ and $\Delta M\simeq\Gamma$  formed by gathering various curves of $R$ for different values of $|b_\mathrm{P}|$ and $\theta$.  The blue and red regions
depict the cases $\Delta M\ll\Gamma$ and $\Delta M\gg\Gamma$, respectively, while orange dashed curves represent the maximum and minimum values for various $k$ in the case $\Delta M\simeq\Gamma$.  To show the trend toward $k \to \infty$, the lower border for large $k$ is drawn for $\theta = \pi - 0.01$, which collapses to $R=0$ as $\theta$ approaches the singular point $\theta = \pi$.}
\label{fig:angularDist}
\end{figure}

Combining all the three cases, we arrive at the overall picture of amplification available by our procedure of postselection as shown in FIG.\ref{fig:angularDist}.  

From this, we learn that, when $\Delta M$ is much smaller than $\Gamma$,  there exists a wide range of $R$ at $k=0$, which will gradually close toward $R=1$ as $\Delta M$ increases compared to $\Gamma$.  In contrast, the behavior of $R$ for $k\to \infty$ approaching $R=2$ is rather consistent and not sensitive to the value $\Delta M$ relative to $\Gamma$.  For a given $k$ which is not too large, the range of amplification is sensitive to the relative relation between the two values, $\Delta M$ and $\Gamma$.

\section{Conclusion and Discussion}\label{sec:conclusion_and_discussion}

In this paper, we present our study on the variation of lifetime of decaying particles (or unstable states) under the application of the weak value amplification technique, which is just to add an extra process of projection measurement called postselection.  By considering a basic two-level system of unstable states dictated by a standard non-Hermitian Hamiltonian, we
found that, compared to the original value, the process of amplification may prolong the decay lifetime up to $2+\sqrt{2}\simeq3.414$ times.   We also observed that, if the postselection filters out the long-lived component in the decaying eigenstates, the lifetime may be shortened arbitrarily (even up to zero) depending on the level of filtering.  

We obtained these results based on a set of assumptions on the nature of the Hamiltonian and the choice of postselection.  These are used for simplifying our analysis, and for the numerical evaluation on the possible range of amplification ({\it i.e.}, variation of lifetime), we considered three extreme cases classified according to the relation between the mass difference $\Delta M$ and the average decay constant $\Gamma$ of the system.  As such, our analysis is not completely general but we believe that these three cases will more or less exhaust the possible range of variation and furnish a reasonably accurate picture on the effect of weak value amplification as may be recognized in FIG.\ref{fig:angularDist}.   In fact, our analysis shows that the amplification factor estimated for the $B$ meson system earlier in \cite{Higashino_2020} may be understood in the context of the range of amplification addressed in this paper.

It would be worth emphasizing that the weak value amplification we discussed here is not quite conventional in that it requires no additional probe system (with some internal degrees of freedom) to capture the weak value and observe the amplification effect.  Rather, the weak value and the associated effect of postselection arise directly in the target system, providing a distinct feature on the lifetime variation compared to other amplification effects observed so far.  In this respect, it is notable that the system exhibiting anomalous time and frequency shifts \cite{Asano_2016} requires no internal degrees of freedom, even though their physical processes appear to be rather different from those of the conventional weak value amplification.  

To summarize, although the observed range of lifetime variation mentioned above may not look large compared from those obtained in the conventional scheme, the range exceeds considerably the eigenvalues of the Hamiltonian and hence may be useful not just for enhancing the accuracy of measurement of lifetime but for delaying the decay process for other purposes of manipulation.  We are hoping to see these to be materialized in the near future.  

Finally, we would like to mention an interesting study of dwelling time of a particle during quantum tunneling \cite{Steinberg_1995, Choi_2013} which bears some similarity with our lifetime of a decaying particle.  These works introduced an operator associated with the dwelling time in a given potential and showed that, by regarding the final position (state) of the particle as the postselection, the transmission and reflection times are defined as the weak values of the dwelling operator for the respective physical processes.  The sum of the transmission time and the reflection time, each weighted with the corresponding probability, then yields the proper dwelling time defined as the expectation value of the operator.  The relation among these quantities follows naturally from the standard identity between the weak values and the expectation value ensured by the unitarity of time evolution.  Despite the similarity,  we also have differences in that our system does not admit a unitary time evolution (since our effective Hamiltonian is non-Hermitian) and that we do not define the conditional lifetime $\tau(\Psi\to \Phi)$ as well as the non-conditional one $\tau(\Psi)$ from a single operator common to them.  Accordingly, one cannot, in principle, expect an analogous relation between $\tau(\Psi\to \Phi)$ and $\tau(\Psi)$ as we have for the dwelling time.  
Nonetheless, it is amusing to observe that, in the linear approximation where the non-unitarity is muted relatively and the weak value arises explicitly as in \eqref{effective_Gamma}, an analogous relation holds (see Appendix B),
\begin{align}
\sum_{k}|\!\braket{\Phi_{k}|\Psi}\!|^{2}\tau(\Psi\to\Phi_{k}) = \tau(\Psi),
\label{reldecays}
\end{align} 
where the summation is over a complete set of postselected states $\ket{\Phi_{k}}$.  This salient property indicates that there may be some deeper connection between the lifetime in particle decay where the weak value arises indirectly and the dwelling time in quantum tunneling where it arises directly.  To uncover such connection, perhaps we need to extend further the scope of the analysis on the effect of postselection {\it per se}, without assuming the involvement of the weak value from the outset, by considering other related issues such as those discussed in \cite{Asano_2016} in a unified context.

\begin{acknowledgments}
We would like to thank Satoshi Higashino, Yosuke Takubo, Takeo Higuchi and Akimasa Ishikawa for useful discussions. This work was supported by JSPS KAKENHI Grant Number 20H01906. 
\end{acknowledgments}


\medskip

\appendix

\section{Conditional lifetime}

Here we provide some formulae needed to derive the conditional lifetime $\tau(\Psi\to \Phi)$ in \eqref{act_lifetime} in a form 
convenient for considering the possible range of variation of the ratio $R$.

First, from \eqref{eq:dm} we obtain the denominator of \eqref{eq:cond-lifetime},
\begin{align}
    \int_{0}^{\infty} dt\, |\langle \Phi|\Psi (t)\rangle|^{2}&=\frac{|c_\mathrm{L}|^{2}}{\Gamma_\mathrm{L}}+\frac{|c_\mathrm{H}|^{2}}{\Gamma_\mathrm{H}}\nonumber\\
    &\quad +\frac{2\Gamma}{\Gamma^{2}+\Delta M^{2}}\mathrm{Re}\left[c_\mathrm{L}^{\ast}c_\mathrm{H}\right]\nonumber\\
    &\quad+\frac{2\Delta M}{\Gamma^{2}+\Delta M^{2}}\mathrm{Im}\left[c_\mathrm{L}^{\ast}c_\mathrm{H}\right],
\label{eq:denominator_of_lifetime}\
\end{align}    
and also the numerator,
\begin{align}
     \int_{0}^{\infty} dt\ t|\langle \Phi|\Psi (t)\rangle|^{2}&=\frac{|c_\mathrm{L}|^{2}}{\Gamma^{2}_\mathrm{L}}+\frac{|c_\mathrm{H}|^{2}}{\Gamma^{2}_\mathrm{H}}\nonumber\\
    &\quad +\frac{-2\Delta M^{2}+2\Gamma^2}{(\Gamma^{2}+\Delta M^{2})^{2}}\mathrm{Re}\left[c_\mathrm{L}^{\ast}c_\mathrm{H}\right]\nonumber\\
    &\quad+\frac{4\Gamma\Delta M}{(\Gamma^{2}+\Delta M^{2})^{2}}\mathrm{Im}\left[c_\mathrm{L}^{\ast}c_\mathrm{H}\right].\label{eq:numerator_of_lifetime}
\end{align}
Plugging these into \eqref{eq:cond-lifetime} yields \eqref{act_lifetime}.

Next, in order to express the obtained conditional lifetime in terms of $b_\mathrm{P}$ and $b_\mathrm{\bar{P}}$, one has to convert $c_\mathrm{L}$ and $c_\mathrm{H}$ into them taking account of their normalization conditions.  To do this, we use \eqref{eq:post-state} to obtain
\begin{align}
    b_\mathrm{L}b_\mathrm{H}^{*}
    &=\frac{|b_\mathrm{P}|^{2}}{4|p|^{2}}-\frac{|b_\mathrm{\bar{P}}|^{2}}{4|q|^{2}}-2i\mathrm{Im}\left[\frac{b_\mathrm{\bar{P}}^{*}b_\mathrm{P}}{4pq^{*}}\right].\label{blbhstar-calc}
\end{align}
Then, by using \eqref{eq:sha}, \eqref{eq:shb} and \eqref{blbhstar-calc}, we find
\begin{align}
    &c_\mathrm{L}^{*}c_\mathrm{H}\nonumber\\
    &=a_\mathrm{L}^{*}a_\mathrm{H}(b_\mathrm{L}b_\mathrm{H}^{*}(1-|\braket{P_\mathrm{L}|P_\mathrm{H}}|^{2})+\braket{P_\mathrm{L}|P_\mathrm{H}})\nonumber\\
    &=\frac{1}{4|p|^{2}}\left[\left(\frac{|b_\mathrm{P}|^{2}}{4|p|^{2}}-\frac{|b_\mathrm{\bar{P}}|^{2}}{4|q|^{2}}\right)(1-(|p|^2-|q|^2)^2)+|p|^2-|q|^2\right]\nonumber\\
    &\qquad-i\frac{1-(|p|^2-|q|^2)^2}{2|p|^{2}}\mathrm{Im}\left[\frac{b_\mathrm{\bar{P}}^{*}b_\mathrm{P}}{4pq^{*}}\right].\label{cont:clchstar}
\end{align}
Now, for the special situation \eqref{restrict} and \eqref{restrictpre}, if we use \eqref{cont:clchstar}
together with \eqref{eq:sha}, \eqref{eq:shb}, we arrive at \eqref{eff-lifetime} and \eqref{fullratio} in the text.

\section{Relation between the conditional and non-conditional lifetimes}

We show that, under the linear approximation, the conventional lifetime $\tau(\Psi)$ given in \eqref{def:taup} and the conditional lifetime $\tau(\Psi\to \Phi)$ given in \eqref{eq:cond-lifetime}
are related in a manner similar to the relation between the expectation value and the weak value.  

To this end, one first recalls \eqref{effective_Gamma} which is valid in the linear approximation.  Then, by summing over a complete set of states $\ket{\Phi_{k}}$ for the postselection with the respective weight factor of the transition probability $|\!\braket{\Phi_{k}|\Psi}\!|^{2}$, one finds
\begin{align}
\!\! \sum_{k}|\!\braket{\Phi_{k}|\Psi}\!|^{2}\tau(\Psi\to\Phi_{k})
&\simeq\frac{1}{\Gamma}+\frac{2g}{\Gamma^{2}}\mathrm{Im}\left[\sum_{k}|\!\braket{\Phi_{k}|\Psi}\!|^{2}A_{\mathrm{w}}\right]\nonumber\\
&=\frac{1}{\Gamma}+\frac{\Delta M}{\Gamma^{2}}\mathrm{Im}[\braket{\Psi|\hat A|\Psi}].
\label{ctone}
\end{align}
With the normalized Hamiltonian $\hat A$ in \eqref{eq:deltah} which is non-Hermitian, one can evaluate its expectation value from \eqref{eq:init-state} to obtain
\begin{align}
\mathrm{Im}[\braket{\Psi|\hat{A}|\Psi}]=\frac{\Delta \Gamma}{2\Delta M}(|a_\mathrm{L}|^{2}-|a_\mathrm{H}|^{2})
+2\mathrm{Im}[a_\mathrm{H}a_\mathrm{L}^{\ast}\! \braket{P_\mathrm{L}|P_\mathrm{H}}]
\end{align}
with $\Delta \Gamma = \Gamma_\mathrm{H}-\Gamma_\mathrm{L}$.  Plugging this into \eqref{ctone} gives
\begin{align}
&\sum_{k}|\!\braket{\Phi_{k}|\Psi}\!|^{2}\tau(\Psi\to\Phi_{k})\nonumber\\
&\quad =\frac{1}{\Gamma}+\frac{\Delta\Gamma}{2\Gamma^{2}}(|a_\mathrm{L}|^{2}-|a_\mathrm{H}|^{2})
+2\frac{\Delta M}{\Gamma^{2}}\mathrm{Im}[a_\mathrm{H}a_\mathrm{L}^{\ast}\braket{P_\mathrm{L}|P_\mathrm{H}}]
\label{summedcl}
\end{align}
up to the linear order of $\Delta M/\Gamma$ and $\Delta \Gamma/\Gamma$.

On the other hand, applying the same linear approximation to \eqref{def:taup}, one finds
\begin{align}
\tau(\Psi) 
    &=\frac{1}{\Gamma}\left(|a_\mathrm{L}|^{2}+|a_\mathrm{H}|^{2} +  2\mathrm{Re}\left[a_\mathrm{H}a_\mathrm{L}^{*}\braket{P_\mathrm{L}|P_\mathrm{H}}\right]\right)
    \nonumber\\
     &\quad +\frac{\Delta\Gamma}{2\Gamma^2}(|a_\mathrm{L}|^{2}-|a_\mathrm{H}|^{2})+2\frac{\Delta M}{\Gamma^{2}}\mathrm{Im}\left[a_\mathrm{H}a_\mathrm{L}^{*}\braket{P_\mathrm{L}|P_\mathrm{H}}\right].
    \label{def:tauptwo}
\end{align}
Using the normalization condition of the preselected state $\ket{\Psi}$ in \eqref{eq:init-state},
\begin{align}
\braket{\Psi|\Psi}=|a_\mathrm{L}|^{2}+|a_\mathrm{H}|^{2}+2\mathrm{Re}[a_\mathrm{H}a_\mathrm{L}^{*}\braket{P_\mathrm{L}|P_\mathrm{H}}]=1,
\end{align}
we observe that the non-conditional lifetime \eqref{def:tauptwo} coincides precisely with the summed over conditional lifetime \eqref{summedcl}.  This shows that the relation \eqref{reldecays} in the text holds for any preselected state $\ket{\Psi}$ at least in the regime of linear approximation.

\bibliography{main}

\begin{thebibliography}{17}%
\makeatletter
\providecommand \@ifxundefined [1]{%
 \@ifx{#1\undefined}
}%
\providecommand \@ifnum [1]{%
 \ifnum #1\expandafter \@firstoftwo
 \else \expandafter \@secondoftwo
 \fi
}%
\providecommand \@ifx [1]{%
 \ifx #1\expandafter \@firstoftwo
 \else \expandafter \@secondoftwo
 \fi
}%
\providecommand \natexlab [1]{#1}%
\providecommand \enquote  [1]{``#1''}%
\providecommand \bibnamefont  [1]{#1}%
\providecommand \bibfnamefont [1]{#1}%
\providecommand \citenamefont [1]{#1}%
\providecommand \href@noop [0]{\@secondoftwo}%
\providecommand \href [0]{\begingroup \@sanitize@url \@href}%
\providecommand \@href[1]{\@@startlink{#1}\@@href}%
\providecommand \@@href[1]{\endgroup#1\@@endlink}%
\providecommand \@sanitize@url [0]{\catcode `\\12\catcode `\$12\catcode
  `\&12\catcode `\#12\catcode `\^12\catcode `\_12\catcode `\%12\relax}%
\providecommand \@@startlink[1]{}%
\providecommand \@@endlink[0]{}%
\providecommand \url  [0]{\begingroup\@sanitize@url \@url }%
\providecommand \@url [1]{\endgroup\@href {#1}{\urlprefix }}%
\providecommand \urlprefix  [0]{URL }%
\providecommand \Eprint [0]{\href }%
\providecommand \doibase [0]{http://dx.doi.org/}%
\providecommand \selectlanguage [0]{\@gobble}%
\providecommand \bibinfo  [0]{\@secondoftwo}%
\providecommand \bibfield  [0]{\@secondoftwo}%
\providecommand \translation [1]{[#1]}%
\providecommand \BibitemOpen [0]{}%
\providecommand \bibitemStop [0]{}%
\providecommand \bibitemNoStop [0]{.\EOS\space}%
\providecommand \EOS [0]{\spacefactor3000\relax}%
\providecommand \BibitemShut  [1]{\csname bibitem#1\endcsname}%
\let\auto@bib@innerbib\@empty
\bibitem [{\citenamefont {Aharonov}\ \emph {et~al.}(1988)\citenamefont
  {Aharonov}, \citenamefont {Albert},\ and\ \citenamefont
  {Vaidman}}]{Aharonov_1988}%
  \BibitemOpen
  \bibfield  {author} {\bibinfo {author} {\bibfnamefont {Y.}~\bibnamefont
  {Aharonov}}, \bibinfo {author} {\bibfnamefont {D.~Z.}\ \bibnamefont
  {Albert}}, \ and\ \bibinfo {author} {\bibfnamefont {L.}~\bibnamefont
  {Vaidman}},\ }\href {\doibase 10.1103/PhysRevLett.60.1351} {\bibfield
  {journal} {\bibinfo  {journal} {Phys. Rev. Lett.}\ }\textbf {\bibinfo
  {volume} {60}},\ \bibinfo {pages} {1351} (\bibinfo {year}
  {1988})}\BibitemShut {NoStop}%
\bibitem [{\citenamefont {Hosten}\ and\ \citenamefont
  {Kwiat}(2008)}]{Hosten_2008}%
  \BibitemOpen
  \bibfield  {author} {\bibinfo {author} {\bibfnamefont {O.}~\bibnamefont
  {Hosten}}\ and\ \bibinfo {author} {\bibfnamefont {P.}~\bibnamefont {Kwiat}},\
  }\href {\doibase 10.1126/science.1152697} {\bibfield  {journal} {\bibinfo
  {journal} {Science}\ }\textbf {\bibinfo {volume} {319}},\ \bibinfo {pages}
  {787} (\bibinfo {year} {2008})}\BibitemShut {NoStop}%
\bibitem [{\citenamefont {Dixon}\ \emph {et~al.}(2009)\citenamefont {Dixon},
  \citenamefont {Starling}, \citenamefont {Jordan},\ and\ \citenamefont
  {Howell}}]{Dixon_2009}%
  \BibitemOpen
  \bibfield  {author} {\bibinfo {author} {\bibfnamefont {P.~B.}\ \bibnamefont
  {Dixon}}, \bibinfo {author} {\bibfnamefont {D.~J.}\ \bibnamefont {Starling}},
  \bibinfo {author} {\bibfnamefont {A.~N.}\ \bibnamefont {Jordan}}, \ and\
  \bibinfo {author} {\bibfnamefont {J.~C.}\ \bibnamefont {Howell}},\ }\href
  {\doibase 10.1103/PhysRevLett.102.173601} {\bibfield  {journal} {\bibinfo
  {journal} {Phys. Rev. Lett.}\ }\textbf {\bibinfo {volume} {102}},\ \bibinfo
  {pages} {173601} (\bibinfo {year} {2009})}\BibitemShut {NoStop}%
\bibitem [{\citenamefont {Li}\ \emph {et~al.}(2018)\citenamefont {Li},
  \citenamefont {Li}, \citenamefont {Zhang}, \citenamefont {Yu}, \citenamefont
  {Lu}, \citenamefont {Liu}, \citenamefont {Zhang},\ and\ \citenamefont
  {Pan}}]{Li_2018}%
  \BibitemOpen
  \bibfield  {author} {\bibinfo {author} {\bibfnamefont {L.}~\bibnamefont
  {Li}}, \bibinfo {author} {\bibfnamefont {Y.}~\bibnamefont {Li}}, \bibinfo
  {author} {\bibfnamefont {Y.-L.}\ \bibnamefont {Zhang}}, \bibinfo {author}
  {\bibfnamefont {S.}~\bibnamefont {Yu}}, \bibinfo {author} {\bibfnamefont
  {C.-Y.}\ \bibnamefont {Lu}}, \bibinfo {author} {\bibfnamefont {N.-L.}\
  \bibnamefont {Liu}}, \bibinfo {author} {\bibfnamefont {J.}~\bibnamefont
  {Zhang}}, \ and\ \bibinfo {author} {\bibfnamefont {J.-W.}\ \bibnamefont
  {Pan}},\ }\href {\doibase 10.1103/PhysRevA.97.033851} {\bibfield  {journal}
  {\bibinfo  {journal} {Phys. Rev. A}\ }\textbf {\bibinfo {volume} {97}},\
  \bibinfo {pages} {033851} (\bibinfo {year} {2018})}\BibitemShut {NoStop}%
\bibitem [{\citenamefont {Denkmayr}\ \emph {et~al.}(2014)\citenamefont
  {Denkmayr}, \citenamefont {Geppert}, \citenamefont {Sponar}, \citenamefont
  {Lemmel}, \citenamefont {Matzkin}, \citenamefont {Tollaksen},\ and\
  \citenamefont {Hasegawa}}]{Denkmayr:2014aa}%
  \BibitemOpen
  \bibfield  {author} {\bibinfo {author} {\bibfnamefont {T.}~\bibnamefont
  {Denkmayr}}, \bibinfo {author} {\bibfnamefont {H.}~\bibnamefont {Geppert}},
  \bibinfo {author} {\bibfnamefont {S.}~\bibnamefont {Sponar}}, \bibinfo
  {author} {\bibfnamefont {H.}~\bibnamefont {Lemmel}}, \bibinfo {author}
  {\bibfnamefont {A.}~\bibnamefont {Matzkin}}, \bibinfo {author} {\bibfnamefont
  {J.}~\bibnamefont {Tollaksen}}, \ and\ \bibinfo {author} {\bibfnamefont
  {Y.}~\bibnamefont {Hasegawa}},\ }\href {https://doi.org/10.1038/ncomms5492}
  {\bibfield  {journal} {\bibinfo  {journal} {Nat. Commun.}\ }\textbf {\bibinfo
  {volume} {5}},\ \bibinfo {pages} {4492 EP } (\bibinfo {year}
  {2014})}\BibitemShut {NoStop}%
\bibitem [{\citenamefont {Wu}\ \emph {et~al.}(2019)\citenamefont {Wu},
  \citenamefont {Zhang}, \citenamefont {Xie}, \citenamefont {Ou}, \citenamefont
  {Chen}, \citenamefont {Wu},\ and\ \citenamefont {Chen}}]{CWWU_2019}%
  \BibitemOpen
  \bibfield  {author} {\bibinfo {author} {\bibfnamefont {C.-W.}\ \bibnamefont
  {Wu}}, \bibinfo {author} {\bibfnamefont {J.}~\bibnamefont {Zhang}}, \bibinfo
  {author} {\bibfnamefont {Y.}~\bibnamefont {Xie}}, \bibinfo {author}
  {\bibfnamefont {B.-Q.}\ \bibnamefont {Ou}}, \bibinfo {author} {\bibfnamefont
  {T.}~\bibnamefont {Chen}}, \bibinfo {author} {\bibfnamefont {W.}~\bibnamefont
  {Wu}}, \ and\ \bibinfo {author} {\bibfnamefont {P.-X.}\ \bibnamefont
  {Chen}},\ }\href {\doibase 10.1103/PhysRevA.100.062111} {\bibfield  {journal}
  {\bibinfo  {journal} {Phys. Rev. A}\ }\textbf {\bibinfo {volume} {100}},\
  \bibinfo {pages} {062111} (\bibinfo {year} {2019})}\BibitemShut {NoStop}%
\bibitem [{\citenamefont {Luo}\ \emph {et~al.}(2019)\citenamefont {Luo},
  \citenamefont {Xie}, \citenamefont {Qiu}, \citenamefont {Zhou}, \citenamefont
  {Liu}, \citenamefont {Li}, \citenamefont {He}, \citenamefont {Zhang},\ and\
  \citenamefont {Sun}}]{Lan_2019}%
  \BibitemOpen
  \bibfield  {author} {\bibinfo {author} {\bibfnamefont {L.}~\bibnamefont
  {Luo}}, \bibinfo {author} {\bibfnamefont {L.}~\bibnamefont {Xie}}, \bibinfo
  {author} {\bibfnamefont {J.}~\bibnamefont {Qiu}}, \bibinfo {author}
  {\bibfnamefont {X.}~\bibnamefont {Zhou}}, \bibinfo {author} {\bibfnamefont
  {X.}~\bibnamefont {Liu}}, \bibinfo {author} {\bibfnamefont {Z.}~\bibnamefont
  {Li}}, \bibinfo {author} {\bibfnamefont {Y.}~\bibnamefont {He}}, \bibinfo
  {author} {\bibfnamefont {Z.}~\bibnamefont {Zhang}}, \ and\ \bibinfo {author}
  {\bibfnamefont {H.}~\bibnamefont {Sun}},\ }\href {\doibase 10.1063/1.5083995}
  {\bibfield  {journal} {\bibinfo  {journal} {{Appl. Phys. Lett.}}\ }\textbf
  {\bibinfo {volume} {114}},\ \bibinfo {pages} {111104} (\bibinfo {year}
  {2019})}\BibitemShut {NoStop}%
\bibitem [{\citenamefont {Ogawa}\ \emph {et~al.}(2020)\citenamefont {Ogawa},
  \citenamefont {Kobayashi},\ and\ \citenamefont {Tomita}}]{Ogawa_2020}%
  \BibitemOpen
  \bibfield  {author} {\bibinfo {author} {\bibfnamefont {K.}~\bibnamefont
  {Ogawa}}, \bibinfo {author} {\bibfnamefont {H.}~\bibnamefont {Kobayashi}}, \
  and\ \bibinfo {author} {\bibfnamefont {A.}~\bibnamefont {Tomita}},\ }\href
  {\doibase 10.1103/PhysRevA.101.042117} {\bibfield  {journal} {\bibinfo
  {journal} {{Phys. Rev. A}}\ }\textbf {\bibinfo {volume} {101}},\ \bibinfo
  {pages} {042117} (\bibinfo {year} {2020})}\BibitemShut {NoStop}%
\bibitem [{\citenamefont {Shomroni}\ \emph {et~al.}(2013)\citenamefont
  {Shomroni}, \citenamefont {Bechler}, \citenamefont {Rosenblum},\ and\
  \citenamefont {Dayan}}]{Shomroni_2013}%
  \BibitemOpen
  \bibfield  {author} {\bibinfo {author} {\bibfnamefont {I.}~\bibnamefont
  {Shomroni}}, \bibinfo {author} {\bibfnamefont {O.}~\bibnamefont {Bechler}},
  \bibinfo {author} {\bibfnamefont {S.}~\bibnamefont {Rosenblum}}, \ and\
  \bibinfo {author} {\bibfnamefont {B.}~\bibnamefont {Dayan}},\ }\href
  {\doibase 10.1103/PhysRevLett.111.023604} {\bibfield  {journal} {\bibinfo
  {journal} {Phys. Rev. Lett.}\ }\textbf {\bibinfo {volume} {111}},\ \bibinfo
  {pages} {023604} (\bibinfo {year} {2013})}\BibitemShut {NoStop}%
\bibitem [{\citenamefont {Higashino}\ \emph {et~al.}(2021)\citenamefont
  {Higashino}, \citenamefont {Mori}, \citenamefont {Takubo}, \citenamefont
  {Higuchi}, \citenamefont {Ishikawa},\ and\ \citenamefont
  {Tsutsui}}]{Higashino_2020}%
  \BibitemOpen
  \bibfield  {author} {\bibinfo {author} {\bibfnamefont {S.}~\bibnamefont
  {Higashino}}, \bibinfo {author} {\bibfnamefont {Y.}~\bibnamefont {Mori}},
  \bibinfo {author} {\bibfnamefont {Y.}~\bibnamefont {Takubo}}, \bibinfo
  {author} {\bibfnamefont {T.}~\bibnamefont {Higuchi}}, \bibinfo {author}
  {\bibfnamefont {A.}~\bibnamefont {Ishikawa}}, \ and\ \bibinfo {author}
  {\bibfnamefont {I.}~\bibnamefont {Tsutsui}},\ }\href {\doibase
  10.1103/PhysRevD.104.033001} {\bibfield  {journal} {\bibinfo  {journal}
  {Phys. Rev. D}\ }\textbf {\bibinfo {volume} {104}},\ \bibinfo {pages}
  {033001} (\bibinfo {year} {2021})}\BibitemShut {NoStop}%
\bibitem [{\citenamefont {Matzkin}(2012)}]{Matzkin_2012}%
  \BibitemOpen
  \bibfield  {author} {\bibinfo {author} {\bibfnamefont {A.}~\bibnamefont
  {Matzkin}},\ }\href {\doibase 10.1088/1751-8113/45/44/444023} {\bibfield
  {journal} {\bibinfo  {journal} {{J. Phys. A: Math. Theor.}}\ }\textbf
  {\bibinfo {volume} {45}},\ \bibinfo {pages} {444023} (\bibinfo {year}
  {2012})}\BibitemShut {NoStop}%
\bibitem [{\citenamefont {Zyla}\ \emph {et~al.}(2020)\citenamefont {Zyla} \emph
  {et~al.}}]{PDG_2020}%
  \BibitemOpen
  \bibfield  {author} {\bibinfo {author} {\bibfnamefont {P.~A.}\ \bibnamefont
  {Zyla}} \emph {et~al.} (\bibinfo {collaboration} {{Particle Data Group}}),\
  }\href {\doibase 10.1093/ptep/ptaa104} {\bibfield  {journal} {\bibinfo
  {journal} {Prog. Theor. Exp. Phys.}\ }\textbf {\bibinfo {volume} {2020}},\
  \bibinfo {pages} {083C01} (\bibinfo {year} {2020})}\BibitemShut {NoStop}%
\bibitem [{\citenamefont {Koike}\ and\ \citenamefont
  {Tanaka}(2011)}]{koiketanaka_11}%
  \BibitemOpen
  \bibfield  {author} {\bibinfo {author} {\bibfnamefont {T.}~\bibnamefont
  {Koike}}\ and\ \bibinfo {author} {\bibfnamefont {S.}~\bibnamefont {Tanaka}},\
  }\href {\doibase 10.1103/PhysRevA.84.062106} {\bibfield  {journal} {\bibinfo
  {journal} {Phys. Rev. A}\ }\textbf {\bibinfo {volume} {84}},\ \bibinfo
  {pages} {062106} (\bibinfo {year} {2011})}\BibitemShut {NoStop}%
\bibitem [{\citenamefont {Mori}\ \emph {et~al.}(2019)\citenamefont {Mori},
  \citenamefont {Lee},\ and\ \citenamefont {Tsutsui}}]{Mori_2019}%
  \BibitemOpen
  \bibfield  {author} {\bibinfo {author} {\bibfnamefont {Y.}~\bibnamefont
  {Mori}}, \bibinfo {author} {\bibfnamefont {J.}~\bibnamefont {Lee}}, \ and\
  \bibinfo {author} {\bibfnamefont {I.}~\bibnamefont {Tsutsui}},\ }\href
  {\doibase 10.1088/1361-6455/ab5200} {\bibfield  {journal} {\bibinfo
  {journal} {J. Phys. B: At. Mol. Opt. Phys.}\ }\textbf {\bibinfo {volume}
  {53}},\ \bibinfo {pages} {015501} (\bibinfo {year} {2019})}\BibitemShut
  {NoStop}%
\bibitem [{\citenamefont {Asano}\ \emph {et~al.}(2016)\citenamefont {Asano},
  \citenamefont {Bliokh}, \citenamefont {Bliokh}, \citenamefont {Kofman},
  \citenamefont {Ikuta}, \citenamefont {Yamamoto}, \citenamefont {Kivshar},
  \citenamefont {Yang}, \citenamefont {Imoto}, \citenamefont {\"{O}zdemir},\
  and\ \citenamefont {Nori}}]{Asano_2016}%
  \BibitemOpen
  \bibfield  {author} {\bibinfo {author} {\bibfnamefont {M.}~\bibnamefont
  {Asano}}, \bibinfo {author} {\bibfnamefont {K.}~\bibnamefont {Bliokh}},
  \bibinfo {author} {\bibfnamefont {Y.}~\bibnamefont {Bliokh}}, \bibinfo
  {author} {\bibfnamefont {A.}~\bibnamefont {Kofman}}, \bibinfo {author}
  {\bibfnamefont {R.}~\bibnamefont {Ikuta}}, \bibinfo {author} {\bibfnamefont
  {T.}~\bibnamefont {Yamamoto}}, \bibinfo {author} {\bibfnamefont
  {Y.}~\bibnamefont {Kivshar}}, \bibinfo {author} {\bibfnamefont
  {L.}~\bibnamefont {Yang}}, \bibinfo {author} {\bibfnamefont {N.}~\bibnamefont
  {Imoto}}, \bibinfo {author} {\bibfnamefont {S.}~\bibnamefont {\"{O}zdemir}},
  \ and\ \bibinfo {author} {\bibfnamefont {F.}~\bibnamefont {Nori}},\ }\href
  {\doibase 10.1038/ncomms13488} {\bibfield  {journal} {\bibinfo  {journal}
  {Nat. Commun.}\ }\textbf {\bibinfo {volume} {7}},\ \bibinfo {pages} {13488}
  (\bibinfo {year} {2016})}\BibitemShut {NoStop}%
\bibitem [{\citenamefont {Steinberg}(1995)}]{Steinberg_1995}%
  \BibitemOpen
  \bibfield  {author} {\bibinfo {author} {\bibfnamefont {A.~M.}\ \bibnamefont
  {Steinberg}},\ }\href {\doibase 10.1103/PhysRevLett.74.2405} {\bibfield
  {journal} {\bibinfo  {journal} {{Phys. Rev. Lett.}}\ }\textbf {\bibinfo
  {volume} {74}},\ \bibinfo {pages} {2405} (\bibinfo {year}
  {1995})}\BibitemShut {NoStop}%
\bibitem [{\citenamefont {Choi}\ and\ \citenamefont
  {Jordan}(2013)}]{Choi_2013}%
  \BibitemOpen
  \bibfield  {author} {\bibinfo {author} {\bibfnamefont {Y.}~\bibnamefont
  {Choi}}\ and\ \bibinfo {author} {\bibfnamefont {A.~N.}\ \bibnamefont
  {Jordan}},\ }\href {\doibase 10.1103/PhysRevA.88.052128} {\bibfield
  {journal} {\bibinfo  {journal} {{Phys. Rev. A}}\ }\textbf {\bibinfo {volume}
  {88}},\ \bibinfo {pages} {052128} (\bibinfo {year} {2013})}\BibitemShut
  {NoStop}%
\end{thebibliography}%

\end{document}